\documentclass[%
 reprint,
 amsmath,amssymb,
 aps,
]{revtex4-2}

\usepackage{graphicx}
\usepackage{dcolumn}
\usepackage{bm}
\usepackage{stmaryrd}
\usepackage{xcolor}
\newcommand{\edit}[1]{#1}
\begin{document}
\preprint{APS/123-QED}

\title{Fourier modal method for gratings with chiral, non-reciprocal and bi-anisotropic materials}

\author{Ilia Smagin}
 \email{Ilia.Smagin@skoltech.ru}
\affiliation{Skolkovo Institute of Science and Technology, Bolshoi Boulevard 30, str 1, Moscow, Russia}

\author{Thomas Weiss}
\affiliation{Institute of Physics, Universitätsplatz 5, University of Graz, and NAWI Graz, 8010 Graz, Austria} 

\author{Sergey Dyakov}
 \email{S.Dyakov@skoltech.ru}
\affiliation{Skolkovo Institute of Science and Technology, Bolshoi Boulevard 30, str 1, Moscow, Russia }

\begin{abstract}
We report an advanced formulation of the Fourier modal method developed for two-dimensionally periodic multilayered structures containing materials with non-zero macroscopic magneto-electric coefficients (also known as coefficients of chirality and bi-anisotropy) represented as arbitrary $3\times 3$ tensors. We consider two numerical schemes for this formulation: \edit{with and without generalized Fourier} factorization rules. For both schemes, we provide explicit expressions for the Fourier tensors of macroscopic material parameters and demonstrate that, in the absence of magneto-electric coupling, they reduce to \edit{the} conventional \edit{factorization rules}. We show that the scheme employing factorization rules facilitates improved convergence, even when the macroscopic chirality coefficient is large. The described formulation represents \edit{a} fast and rigorous technique for theoretical studies of periodic structures with chiral, bi-anisotropic, or non-reciprocal materials in the widely used framework of the Fourier modal method.
\end{abstract} 

\maketitle

\section{Introduction}\label{sec:introduction}

The electrodynamic semi-analytical Fourier modal method (FMM) is one of the most efficient and fastest approaches for theoretical studies of the optical properties of layered periodic structures in photonics. This technique employs a scattering matrix formalism along with the Fourier decomposition of fields in each of the vertically \edit{stacked} homogeneous layers. Since \edit{the} publication of the \edit{first} seminal works \cite{MG1981, TG2002}, this technique has undergone extensive \edit{improvements} and \edit{adaptations}~\cite{fradkin2020nanoparticle, salakhova2021fourier, weiss2009matched}.

\edit{Most} implementations of the FMM are based on constitutive relations that connect the local electric induction with the local electric field and the local magnetic induction with the local magnetic field. Typically, this form of constitutive relation serves as an excellent approximation for local polarization and magnetization, accurately describing experimental observations. However, there is a class of \edit{linear} optical phenomena that cannot be described by these relations. Among these phenomena are circular dichroism --- the difference in the absorption of left- and right\edit{-handed} circularly polarized light by a substance \cite{Johnson1988, Hendry2010, Tang2011}, and optical activity --- the rotation of the polarization plane of linearly polarized light as it travels through the substance \cite{inoue2004chiral, hodgkinson2001inorganic}. Microscopically, these phenomena originate from the broken inversion symmetry (chirality) of molecules that constitute such substances. Circular dichroism and optical activity can be described by the simplest form of constitutive relations only when using a dielectric permittivity \edit{that is} dependent on the polarization state \edit{and the direction or propagation} of light. Although this phenomenological approach can be effective in homogeneous media, it fails in periodic media with intrinsically chiral materials.

Comprehensive research on the optical properties of chiral materials in the past century \cite{fyodorov1976teoriya, lindell1994electromagnetic} can be summarized in constitutive relations containing cross-coupling coefficients between local electric induction and the local magnetic field and local magnetic induction and the local electric field. These cross-coupling coefficients are referred to as macroscopic chirality coefficients, which, along with dielectric permittivity and magnetic permeability, are subject to experimental determination. In natural chiral materials, macroscopic chirality coefficients have relatively small absolute values, typically \edit{below}~$10^{-4}$ . In principle, this allows one to utilize perturbation theory for the generalization of the FMM, since the cross-terms in the constitutive equations are significantly smaller compared to the main terms \cite{almousa2024employing}. However, this method will be inadequate for describing resonant scenarios when the chirality coefficients have poles in their frequency dependences. In addition, artificial chiral materials, such as metamaterials, may also have large chirality coefficients that can make the use of perturbation theory insufficient. Thus, the development of \edit{a} Fourier modal method capable of calculating the scattering matrix of the multilayered structure with chiral substances is of great importance. This is further stimulated by the growing interest of the photonics community in chiral polaritonics, which is in its infancy at the moment.

As shown in~\cite{lindell1994electromagnetic}, constitutive relations with magneto-electric coupling coefficients can describe not only chiral media but also media for which the electromagnetic reciprocity theorem does not hold. Non-reciprocal media require not only broken inversion symmetry, but also broken time-reversal symmetry, which can be achieved by static magnetic field, \edit{nonlinear effects, or time-varying media}~\cite{Floess2016a, nutskii2025complex}.

Chiral and nonreciprocal media are special cases of a more general class of magneto-electric media; the wide variety of optical phenomena associated with these media can be described by constitutive relations with magneto-electric coupling. Although an improvement of the FMM applicable to structures containing magneto-electric but homogeneous layers is quite straightforward \cite{dyakov2024chiral}, the generalization of the FMM to magneto-electric periodic media is much more challenging due to the necessity of using \edit{the correct Fourier} factorization rules \edit{derived by Lifeng Li} for convergence improvement \cite{Li1996, Li1998, Li2003}. \edit{Formulations for one-dimensional periodic systems are available} \edit{ \cite{onishi2011formulation}}. In this paper, we address this problem \edit{for crossed gratings} and formulate the FMM considering macroscopic dielectric permittivity, magnetic permeability, and magneto-electric coefficients in their most general form represented by 3$\times$3 tensors. We will consider two numerical schemes for this formulation: \edit{On the one hand, we completely omit any factorization rules. On the other hand, we apply the correct generalization of
the factorization rules to magneto-electric materials.} For both schemes, we provide explicit expressions for the Fourier tensors of macroscopic material parameters.

\section{Conventional Fourier modal method}
\label{sec:formulation}

In this section, we will formulate \edit{the} basic \edit{concept} of the Fourier modal method in application to structures consisting of materials \edit{without any magneto-electric coupling}. We will also formulate Li's factorization rules, an approach that substantially \edit{improves} the convergence of the numerical scheme.

\subsection{Formulation}
\begin{figure}[t!]
    \centering
    \includegraphics[width=0.8\linewidth]{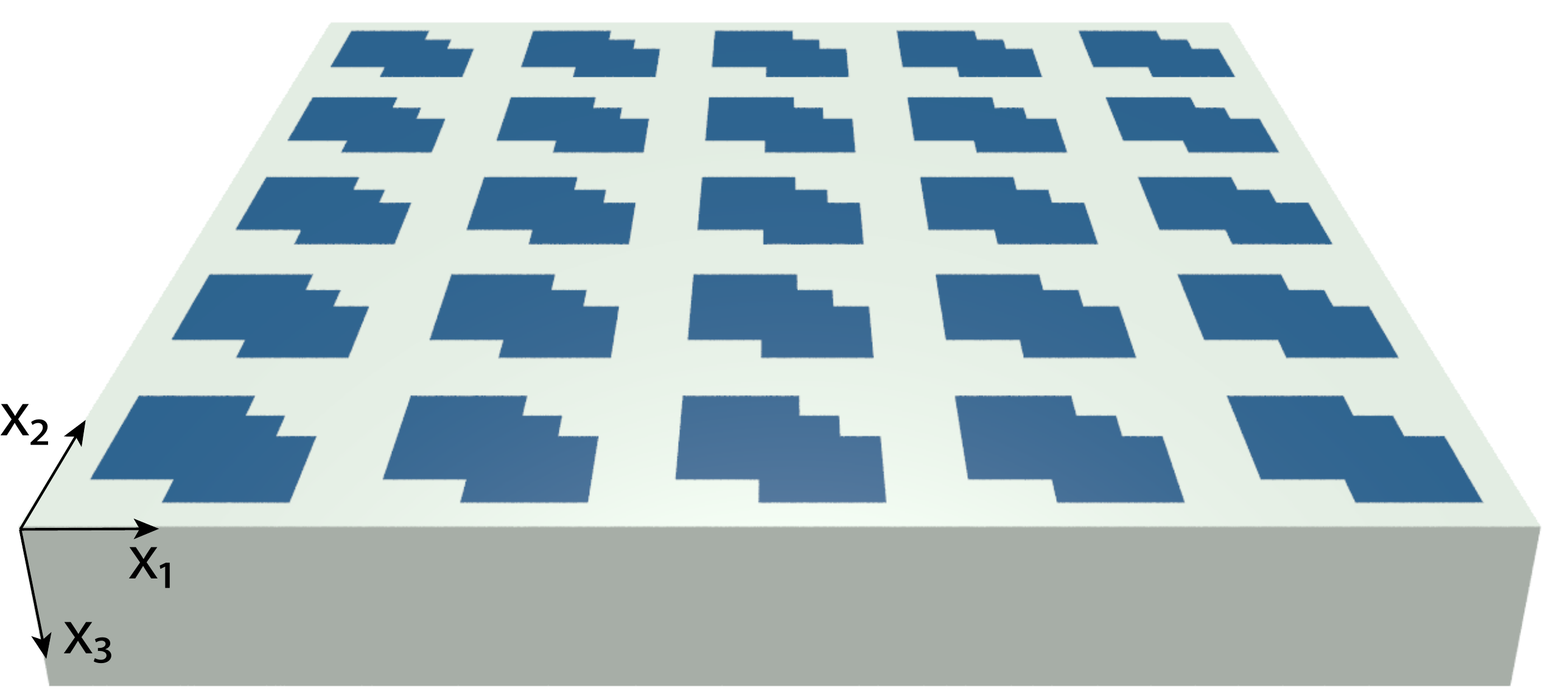}
    \caption{Scheme of a two-dimensional photonic crystal slab. The colors indicate different materials.}
    \label{Layer}
\end{figure}

The principal sketch of a photonic structure to which the FMM can be applied is shown in Fig.~\ref{Layer}. It consists of several layers, each of \edit{them being} periodic along the horizontal plane and homogeneous along the vertical axis. The Maxwell's equations in such a system \edit{can then be solved} using the \edit{iterative} scattering matrix \edit{formalism}, which is based on finding the solutions of an eigenvalue problem in each layer and subsequent connection of solutions of adjacent layers~\cite{TG2002}. \edit{Without the loss of generality,} we \edit{consider a} Cartesian coordinate system, where the \edit{$x^1$ and $x^2$ axes are collinear to basis vectors of the unit cell}, while the coordinate $x^3$ corresponds to the direction along which the layers are \edit{translationally symmetric.}

Let us formulate the eigenvalue problem for a vertically homogeneous layer. \edit{In the absence of magneto-electric coupling, the constitutive material relations are of} form
\begin{align}
        \mathbf{D} &= \underline{\underline{\varepsilon}} \mathbf{E}, &
        \mathbf{B} &= \underline{\underline{\mu}} \mathbf{H},
    \label{eq1vect}
\end{align}
\edit{where the electric and magnetic induction $\mathbf{D}$ and $\mathbf{B}$ are connected to the electric and magnetic fields $\mathbf{E}$ and $\mathbf{H}$ via the macroscopic permittivity and permeability tensors $\underline{\underline{\varepsilon}}$ and $\underline{\underline{\mu}}$, respectively. Using Einstein's sum convention, i.e., a summation is implied whenever there is a pair of identical sub- and superscript, the component-wise form of the constitutive relations reads}
\begin{align}
        \mathrm{D}^{\rho} &= \varepsilon^{\rho\sigma} \mathrm{E}_\sigma, &
        \mathrm{B}^{\rho} &= \mu^{\rho\sigma} \mathrm{H}_\sigma.
    \label{eq1}
\end{align}
\edit{The Greek indices run here over all spatial directions from one to three. In the absence of charge and current densities, the covariant form of Maxwell’s curl equations is}
\begin{align}
        \epsilon^{\rho\sigma\tau}\partial_\sigma \mathrm{E}_\tau &= ik_0\mu^{\rho\sigma} \mathrm{H}_\sigma, &
        \epsilon^{\rho\sigma\tau}\partial_\sigma \mathrm{H}_\tau &= - ik_0\varepsilon^{\rho\sigma} \mathrm{E}_\sigma,
    \label{eq:Maxwell}
\end{align}
where \edit{a time dependence of $\exp(-i\omega t)$ is implied. Furthermore,  $k_0 = \omega/c$ is the vacuum wavenumber, and $\epsilon^{\rho\sigma\tau}$ is the Levi-Civita symbol.}

In periodic photonic structures, $\mathrm{E}_\sigma$ and $\mathrm{H}_\sigma$ satisfy Bloch's theorem:
\begin{align}\label{eq3a}
    \begin{split}
        \mathrm{E}_\tau\left(x^1,x^2,x^3\right) &= \text{e}^{ik_1x^1+ik_2x^2}E_{\tau}(x^1,x^2,x^3),\\
        \mathrm{H}_\tau\left(x^1,x^2,x^3\right) &= \text{e}^{ik_1x^1+ik_2x^2}H_{\tau}(x^1,x^2,x^3),
    \end{split}
\end{align}
\edit{Here, $k_1$ and $k_2$ are the first and second components of the incident wavevector, respectively, while $E_{\tau}$ and $H_{\tau}$ are functions that are periodic in directions one and two, so that they can be Fourier transformed as
\begin{align}\label{eq3b}
    \begin{split}
    E_\tau\left(x^1,x^2,x^3\right) &= \sum\limits_\mathbf{G}\hat{E}_{\tau,\mathbf{G}}\left(x^3\right) \text{e}^{i \mathbf{G}_1 x^1 + i \mathbf{G}_2 x^2},\\
        H_\tau\left(x^1,x^2,x^3\right) &= \sum\limits_\mathbf{G} \hat{H}_{\tau,\mathbf{G}}\left(x^3\right) \text{e}^{i \mathbf{G}_1 x^1 + i \mathbf{G}_1 x^2}.
    \end{split}
\end{align}
Henceforth, the hat indicates quantities that are Fourier transformed. The reciprocal lattice vectors $\mathbf{G}_1$ and $\mathbf{G}_2$ have the explicit form
\begin{align}
    \mathbf{G_1} &= \frac{2\pi}{a_1}m, & \mathbf{G_2} &= \frac{2\pi}{a_2}m,
\end{align}
where $m$ and $n$ are integer values, over which the sum in Eq.~(\ref{eq3b}) actually runs, and $a_1$ and $a_2$ are the periods in the corresponding directions.}

\edit{The next step in the Fourier modal method is to consider the individual layers of the overall system, in which we have a direction of translational symmetry. By eliminating the third components of $\mathbf{E}$ and $\mathbf{H}$ in Eqs.~(\ref{eq:Maxwell}), it is straight-forward to bring the curl equations into the following form~\cite{Li2003}:
\begin{equation}
    -i\partial_3 \mathbb{F}_{||} = \mathcal{M} \mathbb{F}_{||}
\end{equation}
This equation contains supervectors and supermatrices of the form
\begin{align}\label{eq8}
    \mathbb{F}_{||} &= 
    \begin{pmatrix}
        E_1 \\
        E_2 \\
        H_1 \\
        H_2
    \end{pmatrix}, &
    \mathcal{M}&= 
    \begin{pmatrix}
        \mathcal{M}_\mathrm{EE} & \mathcal{M}_\mathrm{EH}\\
        \mathcal{M}_\mathrm{HE} & \mathcal{M}_\mathrm{HH}
    \end{pmatrix}, 
\end{align}
where
\begin{align}
\begin{split}
\label{eq9}
    \mathcal{M}_\mathrm{EE} &= 
    -i\begin{pmatrix}
        -\widetilde{\mu}_{23}\partial_2 \!- \!\partial_1\widetilde{\varepsilon}_{31} & \widetilde{\mu}_{23}\partial_1 \!-\! \partial_1\widetilde{\varepsilon}_{32}\\
        \widetilde{\mu}_{13}\partial_2 \!- \!\partial_2\widetilde{\varepsilon}_{31} & -\widetilde{\mu}_{13}\partial_1 \!- \!\partial_2\widetilde{\varepsilon}_{32}
    \end{pmatrix},
    \\
    \mathcal{M}_\mathrm{EH} &= 
    k_0\begin{pmatrix}
        \widetilde{\mu}_{21} \!- \!\frac{1}{k_0^2}\partial_1\widetilde{\varepsilon}_{33}\partial_2 & \widetilde{\mu}_{22}\! +\! \frac{1}{k_0^2}\partial_1\widetilde{\varepsilon}_{33}\partial_1\\
        -\widetilde{\mu}_{11} \!+\! \frac{1}{k_0^2}\partial_2\widetilde{\varepsilon}_{33}\partial_2 & -\widetilde{\mu}_{12} \!+\! \frac{1}{k_0^2}\partial_2\widetilde{\varepsilon}_{33}\partial_2
    \end{pmatrix},
    \\
    \mathcal{M}_\mathrm{HE} &= k_0
    \begin{pmatrix}
        -\widetilde{\varepsilon}_{21} \!+\! \frac{1}{k_0^2}\partial_1\widetilde{\mu}_{33}\partial_2 & -\widetilde{\varepsilon}_{22} \!- \!\frac{1}{k_0^2}\partial_1\widetilde{\mu}_{33}\partial_1\\
        \widetilde{\varepsilon}_{11} \!+\! \frac{1}{k_0^2}\partial_2\widetilde{\mu}_{33}\partial_2 & \widetilde{\varepsilon}_{12} \!-\! \frac{1}{k_0^2}\partial_2\widetilde{\mu}_{33}\partial_1
    \end{pmatrix},
    \\
    \mathcal{M}_\mathrm{HH} &= 
    -i\begin{pmatrix}
        -\widetilde{\varepsilon}_{23}\partial_2 \!-\! \partial_1\widetilde{\mu}_{31} & \widetilde{\varepsilon}_{23}\partial_1 \!-\! \partial_1\widetilde{\mu}_{32}\\
        \widetilde{\varepsilon}_{13}\partial_2\! - \!\partial_2\widetilde{\mu}_{31} & -\widetilde{\varepsilon}_{13}\partial_1 \!-\! \partial_2\widetilde{\mu}_{32}
    \end{pmatrix}.
    \end{split}
\end{align}
In these formulas,
\begin{align}\label{eq10}
    \widetilde{\varepsilon}^{\rho\sigma}&\equiv l_3^-(\varepsilon^{\rho\sigma}), & \widetilde{\mu}^{\rho\sigma}&\equiv l_3^-(\mu^{\rho\sigma}),    
\end{align}
where the operator $l^-_3$ belongs to an operator class $l^\pm_\tau$, which is defined as follows~\cite{Li2003}:
\begin{gather}\label{eq11}
    l^\pm_\tau(A^{\rho\sigma}) = 
    \begin{cases}
        \left(A^{\tau\tau}\right)^{-1} & \text{for}\ \rho=\sigma=\tau, \\
        \left(A^{\tau\tau}\right)^{-1} A^{\tau\sigma}& \text{for}\  \rho \neq \tau, \sigma = \tau,\\
        A^{\rho\tau}\left(A^{\tau\tau}\right)^{-1}& \text{for}\  \rho = \tau, \sigma \neq \tau,\\
        A^{\rho\sigma}\pm A^{\rho\tau}\left(A^{\tau\tau}\right)^{-1}A^{\tau\sigma}& \text{for}\  \rho \neq \tau,\sigma \neq \tau.\\
    \end{cases} 
\end{gather}
This operator class will be later referred to \textit{Li operator} and can be applied not only to $3\times 3$ matrices but also to $3\times 3$ supermatrices consisting of non-scalar subblocks with appropriate dimensions.}

\edit{Equation~(\ref{eq11}) can be brought into the form of an eigenvalue problem by using the ansatz
\begin{align}
E_\tau(x^1,x^2,x^3)&=\text{e}^{ik_3x^3}\widetilde{E}_\tau(x^1,x^2), \\
H_\tau(x^1,x^2,x^3)&=\text{e}^{ik_3x^3}\widetilde{H}_\tau(x^1,x^2),
\end{align}
which is
\begin{equation}
    k_3\mathbb{F}_{||} = \mathcal{M}\mathbb{F}_{||}.\label{eq:Meig}
\end{equation}
However, equation~(\ref{eq:Meig}) contains a matrix operator on the right-hand side. In order to bring it into an algebraic form, one needs to expand the operator in the $x^1x^2$ plane by suitable basis functions. A natural choice is the discrete Fourier basis that has been already introduced in Eq.~(\ref{eq3b}).}

\edit{Applying the Fourier transform along directions one and two to Eq.~(\ref{eq:Meig}), basically means to replace the derivatives in Eq.~(\ref{eq11}) by diagonal matrices:
\begin{align}
    \partial_1 &\rightarrow iK_1, & \partial_2 &\rightarrow iK_2.
\end{align}
Here $K_1$ and $K_2$ contain $k_1+\mathbf{G}_1$ and $k_2+\mathbf{G}_2$, respectively, on their diagonal. Furthermore, one needs to replace the permittivity and permeability in the arguments of the operator $l_3^-$ in Eq.~(\ref{eq10}) by
\begin{align}
    \hat{\varepsilon}^{\rho\sigma} &= F_1F_2(\varepsilon^{\rho\sigma}), & \hat{\mu}^{\rho\sigma} &= F_1F_2(\mu^{\rho\sigma})
    \label{eq6}
\end{align}
where $F_1$ and $F_2$ are operators that create the matrices for products in Fourier space from the Fourier transform in directions one and two, respectively~\cite{Li2003}.}

\edit{In numerical calculations, the number of Fourier harmonics is truncated to a maximum order $N_g$. Thus, the Fourier-transformed eigenvalue matrix $\hat{\mathcal{M}}$ in Eq.~(\ref{eq:Meig}) becomes a $4N_g\times 4N_g$ matrix, whose eigenvalues and eigenvectors can be obtained by standard algorithms for solving linear eigenvalue problems. The expectation is that for $N_g\rightarrow\infty$, the solutions of the truncated eigenvalue problem converge to the exact solutions.}

The resulting field distribution in each layer is represented as a linear combination of solutions of~\eqref{eq:Meig}. The coefficients of this linear combination are found by solving the final scattering or eigenvalue problem written in terms of the total scattering matrix with dimensionality $4N_g\times4N_g$. The total scattering matrix is found as a sequence of \edit{Redheffer star products}~\cite{redheffer1959inequalities, rumpf2011improved} of the scattering matrices of layers and interfaces:
\begin{equation}\label{eq12}
    \mathcal{S}_{\mathrm{total}} = \mathcal{S}_{N_s,N}\otimes\mathcal{S}_N\otimes\mathcal{S}_{N,N-1}\otimes\dots\otimes\mathcal{S}_{21}\otimes\mathcal{S}_1\otimes\mathcal{S}_{10},
\end{equation}
where $N$ is the number of layers in a multilayered structure, $\mathcal{S}_i$ is the scattering matrix of the $i$-th layer and $\mathcal{S}_{ij}$ is the scattering matrix of the interface between the $i$-th and the $j$-th layer. These matrices can be found as
\begin{align}
    \label{eq13}
    \mathcal{S}_n & = \left[\begin{array}{cc}
        \exp(iK_3^{(n)}d_n)&0\\0 &\exp(iK_3^{(n)}d_n)
    \end{array}\right],\\
    \mathcal{S}_{n,n-1} &= \Xi\{\hat{\mathcal{F}}^{-1}_{n}\hat{\mathcal{F}}_{n-1}^{\phantom{-1}}\}
    \label{eq14}
\end{align}
where $\Xi$ is an operator that converts a transfer matrix into the scattering matrix \cite{saleh2019fundamentals}, ${K}_3^{(n)}$ is a diagonal matrix composed of the eigenvalues of the eigenvalue problem~\eqref{eq:Meig} in the $n$-th layer, and $\hat{\mathcal{F}}_n$ is a matrix containing the corresponding eigenvectors \edit{as columns, such that
\begin{equation}
\hat{\mathcal{F}}_nK_3^{(n)}=\hat{\mathcal{M}}_n\hat{\mathcal{F}}_n,
\end{equation}
provided that $\hat{\mathcal{M}}_n$ is the eigenvalue matrix of the $n$-th layer. Finally, $d_n$ is the thickness of the $n$-th layer.}

The presented numerical scheme of finding the solution of Maxwell's equations in a layer and in the entire multilayered structure is subject to convergence \edit{test} with respect to \edit{$N_g$ as} the number of Fourier harmonics. It is known that in periodic structures with high contrast of dielectric permittivity \edit{in one or several layers}, the matrix elements in \eqref{eq8} may not converge \edit{properly}. Lifeng Li demonstrated that \edit{reasonable convergence can only be obtained when accounting for} special rules when calculating the operator $\hat{\mathcal{M}}$~\cite{Li1996, Li1998, Li2003}.

%%%%%%%%%%%%%%%%%%%%%%%%%%%%%%%%%%%%%
\subsection{Li's factorization rules}
%%%%%%%%%%%%%%%%%%%%%%%%%%%%%%%%%%%%%

The poor convergence in calculation of the matrix elements in $\hat{\mathcal{M}}$ originates mathematically from the Gibb's phenomenon that occurs when approximating a discontinuous function using a \edit{truncated} Fourier \edit{expansion}. In application to Maxwell's equations, this problem \edit{occurs} at vertical material boundaries, where the macroscopic parameters such as dielectric permittivity undergo a jump discontinuity; so do the normal components of electromagnetic covariant vectors. As a result, the convolution of two functions with concurrent jump discontinuities hampers the convergence of the entire numerical scheme. The key point of Li's factorization rules involves using the most appropriate functions (direct or inverse) for the calculation of convolutions in matrix elements. As shown in~\cite{Li1996}, the positive effect for convergence is achieved when rewriting the eigenvalue problem in such a way that it does not contain products of two bounded, piecewise smooth periodic functions that have concurrent but not complementary jump discontinuities \cite{Li1996}. Using this strategy, one can significantly improve the convergence. The final scheme of factorization rules is expressed in terms of Eq.~\eqref{eq8} with the only difference that now the matrices $\widehat{\varepsilon}$ and $\widehat{\mu}$ are calculated using an improved scheme: 
\begin{align}\label{eq15}
    \hat{\varepsilon}^{\rho\sigma} &= L_2L_1(\varepsilon^{\rho\sigma}), & \hat{\mu}^{\rho\sigma} &= L_2L_1(\mu^{\rho\sigma}),    
\end{align}
where 
\begin{equation}\label{eq16}
    L_\tau = l^+_\tau F_\tau l^-_\tau,
\end{equation}
with operators $F_\tau$ and $l^{\pm}_\tau$ defined above.

\edit{A proper proof that the Fourier modal method converges to the exact solutions for $N_g\rightarrow\infty$ is difficult, but numerical experiments indicate that the different schemes converge to the same results at different convergence rates. More specifically, the} different schemes generally give different results even when using equal and large numbers of Fourier harmonics. The scheme set by the expression in Eq.~\eqref{eq6} is the simplest in terms of algebraic expressions, but converges slowly. An advanced scheme represented by Eq.~\eqref{eq15} is more complex for practical implementations but demonstrates \edit{significantly} improved convergence. In addition to these, \edit{Li suggested to change the order of operators $L_1$ and $L_2$ and to averaged over both orders}, which provides improved symmetry in solutions compared to scheme~\eqref{eq15}, although it sacrifices energy balance. \edit{Selectively picking out elements of orders $L_1L_2$ and $L_2L_1$ provides a symmetric formulation that also accounts for energy balance, but it can only be applied to special permittivity and permeability tensors~\cite{weiss2009matched}.} In the implementation of the FMM, it is necessary to choose a scheme that is most suitable for a particular situation. Yet, for further advancements, this method requires modifications when addressing materials that exhibit non-zero macroscopic magneto-electric coefficients.

%%%%%%%%%%%%%%%%%%%%%%%%%%%%%%%%%%%%%%%%%%%%%%%%%%%%%%%%%%%%%%%%%%%%%%%%%%%%%%%%%%%%%
\section{Fourier modal method for magneto-electric media}
%%%%%%%%%%%%%%%%%%%%%%%%%%%%%%%%%%%%%%%%%%%%%%%%%%%%%%%%%%%%%%%%%%%%%%%%%%%%%%%%%%%%%

Extensive research on optical activity in the past century \cite{fyodorov1976teoriya, lindell1994electromagnetic, simovski2018composite} can be summarized in the following constitutive relations, which are applicable to a general magneto-electric bi-anisotropic medium:
\begin{gather}\label{eq18}
    \begin{cases}
        \mathrm{D}^\rho = \varepsilon^{\rho\sigma} \mathrm{E}_\sigma + \edit{\xi}^{\rho\sigma} \mathrm{H}_\sigma, \\
        \mathrm{B}^{\rho} = \edit{\zeta}^{\rho\sigma} \mathrm{E}_\sigma + \mu^{\rho\sigma} \mathrm{H}_\sigma,
    \end{cases}
\end{gather}
where $\edit{\xi^{\rho\sigma}}$ and $\edit{\zeta^{\rho\sigma}}$ are \edit{the elements of the} local macroscopic magneto-electric tensors.

For the isotropic case, \edit{where $\xi^{\rho\sigma}=\xi\delta^{\rho\sigma}$ and $\zeta^{\rho\sigma}=\zeta\delta^{\rho\sigma}$,} it can be shown that when $\edit{\xi = -\zeta}$, equations~\eqref{eq18} describe a chiral reciprocal medium. In the literature \edit{(see~\cite{lindell1994electromagnetic} and references therein)}, this medium is referred to as a Pasteur medium, and $i\edit{\xi}$ carries the meaning of the chirality parameter, also known as the Pasteur parameter. A solution of randomly oriented chiral molecules with the same chirality is an example of such a Pasteur medium. Conversely, the case of $\edit{\xi = \zeta}$ corresponds to a non-chiral non-reciprocal medium termed a Tellegen medium, where $\edit{\xi}$ is known as a non-reciprocity parameter or Tellegen response. It should be noted that Tellegen media are much rarer than Pasteur media, with only a limited number of realizations of the former being reported \cite{liu2025photonic, yang2025gigantic, safaei2024optical, shaposhnikov2023emergent}.

It should be kept in mind that the parameters $\edit{\xi}$ and $\edit{\zeta}$ cannot take just any value; they are subject to constraints arising from the condition that the local energy dissipation rate must not be negative \edit{in passive media}~\cite{seidov2025unbounded}. Specifically, for the dispersionless thermodynamically stable Pasteur medium, the chirality parameter is limited by $|\mathrm{Im}\{i\edit{\xi}\}| \le \sqrt{\mathrm{Im}\{\varepsilon\}\mathrm{Im}\{\mu\}}$. While the real part of the chirality parameter $i\edit{\xi}$ is unrestricted, the medium's properties change significantly beyond the condition $|i\edit{\xi}| \le \sqrt{\varepsilon\mu}$ \cite{lindell1994electromagnetic}. For a dispersionless thermodynamically stable Tellegen medium, the non-reciprocity parameter must satisfy the condition $|\edit{\xi}| \le \sqrt{\varepsilon\mu}$ \cite{lindell1994electromagnetic, brown1968upper, seidov2025unbounded}. In media with material resonances, the expression for the dissipation rate is modified, which eventually changes the above restrictions for the parameters \cite{seidov2025unbounded}. 

In the following, we consider the most general magneto-electric bi-anisotropic media with arbitrary tensors $\underline{\underline{\varepsilon}}$, $\underline{\underline{\mu}}$, $\edit{\underline{\underline{\xi}}}$, and $\edit{\underline{\underline{\zeta}}}$, assuming only that the local energy dissipation rate is not negative, and the medium is in thermodynamic equilibrium.

% ----------------------------------------------------------------------------------------------
\subsection{Factorizing general linear materials in the FMM}
% ----------------------------------------------------------------------------------------------
\edit{
Let us rewrite the general material relations for linear response~\eqref{eq18} in a compact notation:
\begin{equation}
	\mathbb{G} = \mathcal{P}\mathbb{F}.\label{eq:Mmat}
\end{equation}
\edit{Here, we have introduced supervectors $\mathbb{F}$ and $\mathbb{G}$ as well as the supermatrix $\mathcal{P}$ as}
\begin{align}
	\mathbb{G}&=\left(\begin{array}{c} \mathbf{D} \\\mathbf{B} \end{array}\right), & \mathbb{F}&=\left(\begin{array}{c} \mathbf{E} \\\mathbf{H} \end{array}\right), & \mathcal{P} = \left(\begin{array}{cc} \varepsilon & \xi \\ \zeta & \mu\end{array}\right).
\end{align}
\edit{By defining a new sum convention, where a sum is carried from one to six out over pairs of identical Greek sub- and superscripts, equation~\eqref{eq:Mmat} can be represented as}
\begin{equation}
	\mathbb{G}^\alpha = \mathcal{P}^{\alpha\beta}\mathbb{F}_\beta.\label{eq:Mcomp}
\end{equation}
\edit{Furthermore, we may introduce two subsets of indices, namely $S_\text{E}$ and $S_\text{H}$, together with the following notation:
\begin{equation}
    [\alpha] = \begin{cases} \alpha & \text{for}\ \alpha\in S_\text{E}=\{1,2,3\}, \\
    \alpha-3 & \text{for}\ \alpha\in S_\text{H}=\{4,5,6\}.\end{cases}
\end{equation}}

\edit{Since it is our goal to carry out Fourier transforms in directions $1$ and $2$, it makes sense to introduce subsets of the form
\begin{align}
    S_{[\alpha]}&=\{[\alpha],[\alpha]+3\}, & \overline{S}_{[\alpha]} &= \{1,2,3,4,5,6\}\setminus S_{[\alpha]},
\end{align}
e.g., $S_1 = \{1,4\}$ and $\overline{S}_{1}=\{2,3,5,6\}$. Henceforth, we will use Latin indices $n$ for subsets $S_{[\alpha]}$, while indices from subsets $\overline{S}_{[n]}$ will be indicated by a bar on top, i.e., $\overline{n}$.}
Thus, we can pick out such subsets \edit{in Eq.~(\ref{eq:Mcomp})}:
\begin{equation}
	\mathbb{G}^n =\mathcal{P}^{nm}\mathbb{F}_m + \mathcal{P}^{n\overline{m}}\mathbb{F}_{\overline{m}}.
\end{equation}
\edit{This equation can be reformulated as
\begin{equation}
    \mathcal{P}^{nm}\mathbb{F}_m = \mathbb{G}^n-\mathcal{P}^{n\overline{m}}\mathbb{F}_{\overline{m}}.
\end{equation}}
Since $n$ and $m$ run over the two same indices \edit{in a given subset $S_{[\alpha]}$}, $\mathcal{P}^{nm}$ are elements of a $2\times 2$ matrix in real space, while they are infinite matrices as subblocks of a $2\times2$ supermatrix in Fourier space. For block-type matrices of the form
\begin{equation}
	M=\left(\begin{array}{cc} A & B \\ C & D \end{array}\right),
\end{equation}
the inverse matrix can be calculated as
\begin{widetext}
\begin{equation}
	M^{-1}=\left[\begin{array}{cc} (A-BD^{-1}C)^{-1} & -(A-BD^{-1}C)^{-1}BD^{-1} \\ -(D-CA^{-1}B)^{-1}CA^{-1} & (D-CA^{-1}B)^{-1} \end{array}\right],\label{eq:Minv}
\end{equation}
\end{widetext}
provided that \edit{subblocks} $A$ and $D$ can be inverted. Let us therefore define a matrix $\mathcal{Q}$ \edit{as
\begin{equation}
    \mathcal{Q}_{\alpha\beta}=I_{[\gamma]}(\mathcal{P}^{\alpha\beta})=\begin{cases} \mathcal{Q}_{nm} & \text{for}\ \alpha,\beta\in S_{[\gamma]},\\
    0 & \text{otherwise},\end{cases}\label{eq:Qdef}
\end{equation}
where $\mathcal{Q}_{nm}$ is calculated via Eq.~(\ref{eq:Minv})} such that
\begin{equation}
	\mathcal{Q}_{pn}\mathcal{P}^{nm} = \delta_p^m.
\end{equation}
Thus,
\begin{equation}
	\mathbb{F}_m = \mathcal{Q}_{mn}\mathbb{G}^n-\mathcal{Q}_{mn}\mathcal{P}^{n\overline{m}}\mathbb{F}_{\overline{m}}.\label{eq:TrafoFm}
\end{equation}
This equation can be Fourier-transformed in the direction \edit{$[m]$, since it contains no products of quantities with concurrent jump discontinuities.}

For the other components, one starts with
\begin{equation}
	\mathbb{G}^{\overline{n}} =\mathcal{P}^{\overline{n}m}\mathbb{F}_m + \mathcal{P}^{\overline{n}\overline{m}}\mathbb{F}_{\overline{m}}.
\end{equation}
Then, one can use Eq.~(\ref{eq:TrafoFm}) to obtain:
\begin{equation}
	\begin{split}
	\mathbb{G}^{\overline{n}} &=\mathcal{P}^{\overline{n}m}(\mathcal{Q}_{mn}\mathbb{G}^n-\mathcal{Q}_{mn}\mathcal{P}^{n\overline{m}}\mathbb{F}_{\overline{m}}) + \mathcal{P}^{\overline{n}\overline{m}}\mathbb{F}_{\overline{m}} \\
	&= \mathcal{P}^{\overline{n}m}\mathcal{Q}_{mn}\mathbb{G}^n + (\mathcal{P}^{\overline{n}\overline{m}}-\mathcal{P}^{\overline{n}m}\mathcal{Q}_{mn}\mathcal{P}^{n\overline{m}})\mathbb{F}_{\overline{m}}.\label{eq:TrafoGn}
	\end{split}
\end{equation}
Again, this equation can be Fourier transformed in direction $[m]$.

Basically, we may introduce a \edit{generalized class of} Li operators $L_{[\gamma]}^\pm$ such that
\begin{equation}
L_{[\gamma]}^\pm(\mathcal{P}^{\alpha\beta})=\begin{cases}Q_{\alpha\beta} & \text{for}\ \alpha,\beta\in S_{[\gamma]},\\
		\mathcal{L}^{\alpha}_\beta & \text{for}\ \alpha\in \overline{S}_{[\gamma]}  \land \beta\in S_{[\gamma]},\\
			\mathcal{R}_{\alpha}^{\beta} & \text{for}\ \alpha\in S_{[\gamma]}  \land\beta\in \overline{S}_{[\gamma]},\\\mathcal{W}^{\alpha\beta}_\pm & \text{for}\ \alpha,\beta\in \overline{S}_{[\gamma]},
	\end{cases}\label{eq:LiOp}
\end{equation}
\edit{where $Q_{\alpha\beta}=I_{[\gamma]}(\mathcal{P}^{\alpha\beta})$ according to Eq.~(\ref{eq:Qdef}), and
\begin{subequations}
\begin{align}
    \mathcal{L}^{\alpha}_\beta &= \mathcal{P}^{\alpha\gamma}Q_{\gamma\beta}, \\
    \mathcal{R}_{\alpha}^{\beta}&=Q_{\alpha\gamma}\mathcal{P}^{\gamma\beta}, \\
    \mathcal{W}^{\alpha\beta}_\pm&=\mathcal{P}^{\alpha\beta} \pm  \mathcal{P}^{\alpha\gamma}Q_{\gamma\delta}\mathcal{P}^{\delta\beta}.
\end{align}
\end{subequations}}
Hence, we may safely Fourier transform $\hat{\mathcal{L}}_{[\gamma]}^- (\mathcal{P}^{\alpha\beta})$ in directions $[\gamma]$.

We still need to revert Eqs.~(\ref{eq:TrafoFm}) and~(\ref{eq:TrafoGn}) \edit{after carrying out the Fourier transform. Using again the hat to indicate Fourier-transformed quantities, we find:}
\begin{align}
	\hat{\mathbb{F}}_m &= \hat{\mathcal{Q}}_{mn}\hat{\mathbb{G}}^n-\hat{\mathcal{R}}_{m}^{\overline{m}}\hat{\mathbb{F}}_{\overline{m}} , \\
	\hat{\mathbb{G}}^{\overline{n}} &= \hat{\mathcal{L}}^{\overline{n}}_n \hat{\mathbb{G}}^n + \hat{\mathcal{W}}_-^{\overline{n}\overline{m}}\hat{\mathbb{F}}_{\overline{m}}.\label{eq:GLW}
\end{align}
The first equation can be easily inverted to yield
\begin{equation}
	\hat{\mathbb{G}}^n = \hat{\mathcal{V}}^{nm}\hat{\mathbb{F}}_m + \hat{\mathcal{V}}^{nm} \hat{\mathcal{R}}_{m}^{\overline{m}}\hat{\mathbb{F}}_{\overline{m}}.\label{eq:backG}
\end{equation}
Here, we have introduced the Fourier-transformed inverse matrix \edit{$\hat{\mathcal{V}}^{\alpha\beta} = I_{[\gamma]}(\hat{\mathcal{Q}}_{\alpha\beta})$} such that
\begin{equation}
	\hat{\mathcal{Q}}_{pn}\hat{\mathcal{V}}^{nm} = \hat{\delta}_p^m,
\end{equation}
which can be also constructed via Eq.~(\ref{eq:Minv}). Inserting Eq.~(\ref{eq:backG}) into Eq.~(\ref{eq:GLW}), we obtain:
\begin{equation}
	\hat{\mathbb{G}}^{\overline{n}} = \hat{\mathcal{L}}^{\overline{n}}_n \hat{\mathcal{V}}^{nm}\hat{\mathbb{F}}_m+(\hat{\mathcal{W}}_-^{\overline{n}\overline{m}} + \hat{\mathcal{L}}^{\overline{n}}_n \hat{\mathcal{V}}^{nm} \hat{\mathcal{R}}_{m}^{\overline{m}})\hat{\mathbb{F}}_{\overline{m}}.
\end{equation}

Thus, we observe that the Fourier transform of Eq.~(\ref{eq:Mcomp}) along direction $[\gamma]$ with the correct Fourier factorization becomes
\begin{equation}
	\hat{\mathbb{G}}^\alpha = L^+_{[\gamma]}F_{[\gamma]}L^-_{[\gamma]}(\mathcal{P}^{\alpha\beta})\hat{\mathbb{F}}_\beta.
\end{equation}
}

% ---------------------------------------------------------------------------------------
\subsection{Eigenvalue problem for a layer with magneto-electric bi-anisotropic materials}
% ---------------------------------------------------------------------------------------
\edit{
To write Maxwell's equations in a compact form, let us introduce a generalized Levi-Civita symbol as
\begin{equation}
	\mathcal{E}^{\alpha\beta\gamma}=\begin{cases} -\epsilon^{\alpha[\beta][\gamma]} & \text{for}\ \alpha\in S_\text{E} \land \beta,\gamma\in S_\text{H},\\
		\epsilon^{[\alpha]\beta\gamma} & \text{for}\ \alpha\in S_\text{H} \land \beta,\gamma\in S_\text{E},\\
		0 & \text{otherwise},\end{cases}
\end{equation}
where $\epsilon^{\alpha\beta\gamma}$ is the conventional Levi-Civita symbol. Together with the operator
\begin{equation}
    \mathcal{D}_\alpha = \partial_{[\alpha]},
\end{equation}
Maxwell's curl equations can be written in frequency domain as
\begin{equation}
	\mathcal{E}^{\alpha\beta\gamma}\mathcal{D}_\beta \mathbb{F}_\gamma = ik_0\mathcal{P}^{\alpha\beta}\mathbb{F}_\beta,
\end{equation}
or in an equivalent form
\begin{equation}
	 \mathbb{G}^\alpha = \frac{1}{ik_0} \mathcal{E}^{\alpha\beta\gamma}\mathcal{D}_\beta \mathbb{F}_\gamma .\label{eq:Gbeta}
\end{equation}

By defining a covariant generalized Levi-Civita symbol with subscript indices that obeys the relation
\begin{equation}
	\mathcal{E}_{\alpha\sigma\tau}\mathcal{E}^{\alpha\beta\gamma}=\delta_\sigma^\beta\delta_\tau^\gamma - \delta_\sigma^\gamma\delta_\tau^\beta,
\end{equation}
which follows from the corresponding relations for the \edit{standard} Levi-Civita symbol, Maxwell's equations can be reformulated as
\begin{equation}
	\mathcal{D}_\sigma\mathbb{F}_\tau - \mathcal{D}_\tau\mathbb{F}_\sigma = ik_0\mathcal{E}_{\alpha\sigma\tau} \mathcal{P}^{\alpha\beta}\mathbb{F}_\beta.
\end{equation}
\edit{Consider now $m\in S_{[\gamma]}$.} This results in
\begin{align}
    \begin{split}
	\mathcal{D}_m\mathbb{F}_{\overline{m}} - \mathcal{D}_{\overline{m}}\mathbb{F}_m = ik_0\mathcal{E}_{\alpha m\overline{m}} \mathcal{P}^{\alpha\beta}\mathbb{F}_\beta = \\
    = ik_0\mathcal{E}_{\overline{n} m\overline{m}} (\mathcal{P}^{\overline{n}p}\mathbb{F}_p+\mathcal{P}^{\overline{n}\overline{p}}\mathbb{F}_{\overline{p}}).\label{eq:partialFm}
    \end{split}
\end{align}
Next, we replace $\mathbb{G}^n$ in Eq.~(\ref{eq:TrafoFm}) by Eq.~(\ref{eq:Gbeta}):
\begin{align}
    \begin{split}
	\mathbb{F}_m &= \mathcal{Q}_{mn}(\mathbb{G}^n-\mathcal{P}^{n\overline{m}}\mathbb{F}_{\overline{m}})=\\
    &=\frac{1}{ik_0}\mathcal{Q}_{mn}( \mathcal{E}^{n\overline{p}\overline{m}}\mathcal{D}_{\overline{p}}-ik_0\mathcal{P}^{n\overline{m}})\mathbb{F}_{\overline{m}}
    \end{split}
\end{align}
Inserting this relation into Eq.~(\ref{eq:partialFm}) results in
\begin{equation}
	\begin{split}
\mathcal{D}_m\mathbb{F}_{\overline{m}}=&\bigg[ik_0\mathcal{E}_{\overline{n} m\overline{m}}(\mathcal{P}^{\overline{n}\overline{q}}-\mathcal{P}^{\overline{n}p}\mathcal{Q}_{pr}\mathcal{P}^{r\overline{q}}) \\
&+ \mathcal{D}_{\overline{m}}\frac{1}{ik_0}\mathcal{Q}_{mn} \mathcal{E}^{n\overline{p}\overline{q}}\mathcal{D}_{\overline{p}} \bigg.  \\
	&-\bigg. \mathcal{D}_{\overline{m}}\mathcal{Q}_{mn}\mathcal{P}^{n\overline{q}} + \mathcal{E}_{\overline{n} m\overline{m}}\mathcal{P}^{\overline{n}p}\mathcal{Q}_{pr} \mathcal{E}^{r\overline{r}\overline{q}}\mathcal{D}_{\overline{r}} \bigg]\mathbb{F}_{\overline{q}}.
	\end{split}
\end{equation}
This can be brought into the compact form
\begin{equation}
	-i\mathcal{D}_m\mathbb{F}_{\overline{m}}=\mathcal{M}_{m\overline{m}}^{\overline{q}}\mathbb{F}_{\overline{q}}.
\end{equation}
\edit{For $m\in S_3$,} $\mathcal{M}$ represents the principal matrix of the Fourier modal method for the considered layer, analogous to \edit{Eq.}~\eqref{eq11} in the absence of the local macroscopic magneto-electric tensors $\xi$ and $\zeta$. The \edit{elements of the} matrix $\mathcal{M}$ \edit{are}
\begin{widetext}
\begin{equation}
	\mathcal{M}_{\alpha\beta}^\gamma = \begin{cases}
		i\mathcal{D}_\beta \delta_{\alpha\nu}\hat{L}_3^-( \mathcal{P}^{\nu\gamma}) -i\mathcal{E}_{\nu\alpha\beta}\mathcal{E}^{\rho\sigma\gamma} \delta_{\tau\rho}\hat{L}_3^-(\mathcal{P}^{\nu\tau})\mathcal{D}_\sigma & \text{for}\ (\beta,\gamma\in S_\text{E})\lor(\beta,\gamma\in S_\text{H}), \\
		k_0 \mathcal{E}_{\sigma \alpha\beta}\hat{L}_3^-( \mathcal{P}^{\sigma\gamma})-k_0^{-1}\mathcal{D}_\beta \delta_{\rho\sigma}\mathcal{E}^{\rho \tau\gamma}\hat{L}_3^-( \mathcal{P}^{\alpha\sigma})\mathcal{D}_\tau & \text{otherwise}.
	\end{cases}
\end{equation}
\end{widetext}
}
% ---------------------------------------------------------------------------------------
\section{Convergence comparison}
% ---------------------------------------------------------------------------------------

\begin{figure}[t!]
    \centering
    \includegraphics[width=0.7\linewidth]{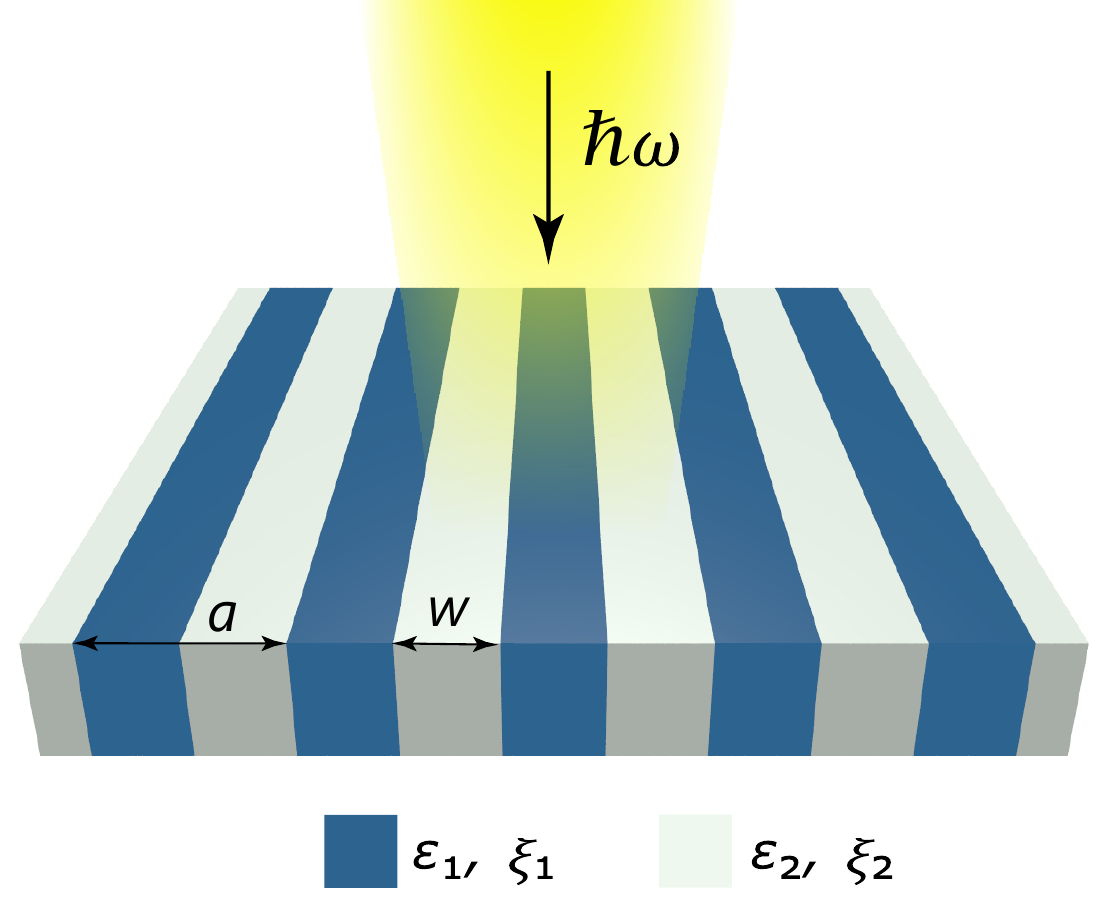}
    \caption{One-dimensional grating for the demonstration of convergence. The colors indicate different materials.}
    \label{1DChiralStructure}
\end{figure}
\begin{figure*}[t!]
    \includegraphics[width=0.8\linewidth]{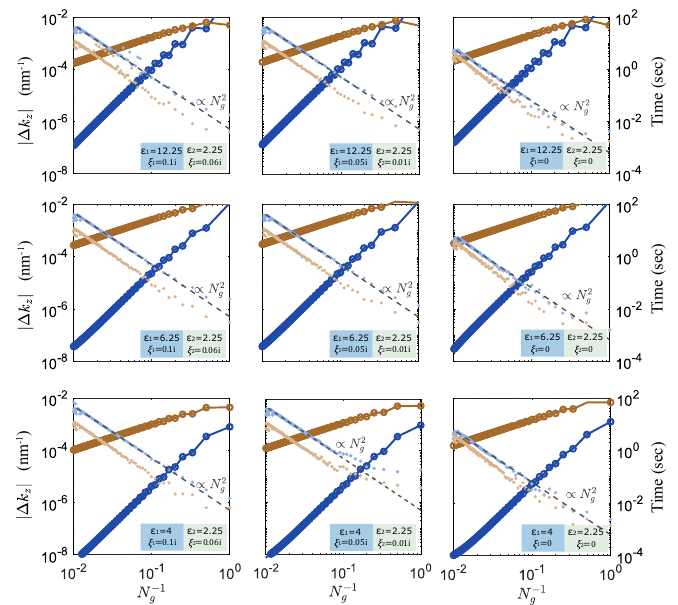}
    \caption{Convergence of the two numerical schemes to the analytical solution in a double logarithmic scale for different values of the refractive indices and chirality coefficients. The \edit{brown} and blue curves represent Scheme 1 (without factorization rules) and Scheme 2 (with factorization rules), respectively. $|\Delta k_z|$ denotes the difference between the $k_z$ value calculated by numerical Schemes 1 and 2 and the exact value, obtained analytically, \edit{as a function of inverse truncation order $N_g^{-1}$}. The computation time for the main matrix $\mathcal{M}$ is shown by the \edit{light brown and light blue} dots for each number of harmonics. \edit{Furthermore,} $\hbar\omega = 1320$~meV, $k_1 = k_2 = 0$. For calculations, we used a 14-core 12th Gen Intel(R) Core(TM) i7-12700H processor, with 16~GB of RAM. }
    \label{Convergence Kz 0}
\end{figure*}

\edit{We consider two numerical schemes. Scheme 1 does not employ \edit{any} factorization rules, while Scheme 2 is formulated using factorization rules for all four tensors $\underline{\underline{\varepsilon}}$, $\underline{\underline{\xi}}$, $\underline{\underline{\zeta}}$, and $\underline{\underline{\mu}}$.} To illustrate the convergence of the considered numerical schemes, we use the example of a one-dimensional photonic crystal slab shown in Fig.~\ref{1DChiralStructure}. The structure consists of alternating stripes of two different materials with a period of $a = 500$~nm. We consider the materials to be isotropic and reciprocal, so that $\edit{\xi}$ and $\edit{\zeta}$ are scalars and $\edit{\xi=-\zeta}$. In the following, we consider different pairs of $\varepsilon$ and $\edit{\xi}$ to see the impact of the material contrast on the convergence (see Fig.~\ref{Convergence Kz 0}). Calculations are made for the photon energy $\hbar\omega = 1320$~meV and zero in-plane wavevector ($k_1 = k_2 = 0$).

\begin{figure*}[t!]
\centering
    \includegraphics[width=0.8\linewidth]{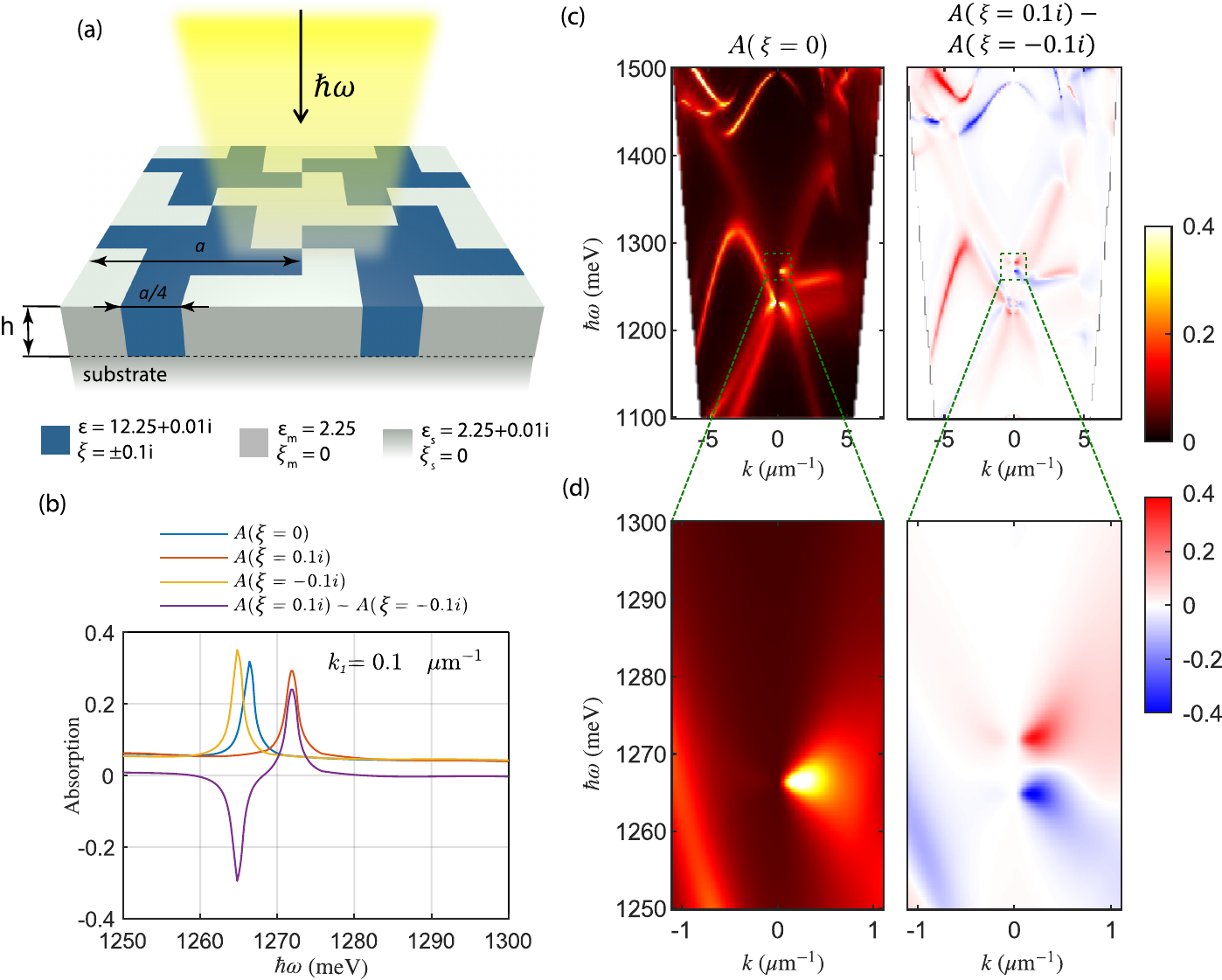}
    \caption{(a) Schematic illustration of the metasurface, comprising a periodic array of chiral elements on a substrate. The blue and grey colors represent materials with distinct dielectric permittivities $\varepsilon$ and chirality parameters $\edit{\xi}$ (b) Absorption spectra calculated in $p$-polarization at $k_1=0.1$~$\mu$m$^{-1}$. \edit{Left panels in} (c) and (d): Photon energy and in-plane wavevector dependence of the absorption coefficient calculated for the non-chiral case with $\edit{\xi=\zeta}^*=0$. \edit{Right panels in} (c) and (d): The difference of the absorption spectra calculated for different signs of chiral coefficients. In (c) and (d) the positive and negative values of $k$ correspond to the $\Gamma-X$ and $\Gamma-M$ directions in reciprocal space. The calculation was performed by selecting 11 \edit{Fourier} harmonics \edit{symmetrically around $\mathbf{G}_1=\mathbf{G}_2=0$ in both reciprocal directions}.}
    \label{AbsorptionSpectra}
\end{figure*}

To investigate the convergence of the numerical schemes, we solve the eigenvalue problem~\eqref{eq:Meig} for each number of Fourier harmonics $N_g$ in the range between 1 and 100, where the matrix $\hat{\mathcal{M}}$ is calculated either by Scheme~1 or Scheme~2. Then, from each solution, we \edit{select the} eigenvalue $k_3$ \edit{with the highest value of} $\text{Re}(k_z)$ and compare this approximate quantity with the exact solution obtained analytically (see Appendix~\ref{appA} for details on finding the exact value of $k_3$).

In Fig.~\ref{Convergence Kz 0}, we plot the absolute value of the deviation of the approximate $k_3$ from the exact value, $|\Delta k_3|$, as a function of the number of Fourier harmonics $N_g$. One can see from Fig.~\ref{Convergence Kz 0} that in the considered range of $N_g$, in all graphs, the deviation $|\Delta k_3|$ decreases as $N_g^{-p}$, where $p = 1$ for Scheme~1 and varies between 1 and 3 for Scheme~2. One can also observe that for all pairs $n$ and $\chi$ Scheme~2 converges faster than Scheme~1. 

In Fig.~\ref{Convergence Kz 0}, we also plot the computation time required to solve the eigenvalue problem~\eqref{eq:Meig} as a function of the number of Fourier harmonics, $N_g$. This time represents the total duration for the numerical solver to compute the eigenvalues for a given matrix size $4N_g\times4N_g$. \edit{As one can see} from the resulting curve, in the considered range of $N_g$ the computation time scales with the square of the number of Fourier harmonics (that is, $t \propto N_g^{2}$).

\section{Numerical example}
As a numerical demonstration, we consider a metasurface consisting of a square lattice of geometrically chiral elements embedded in a matrix. The structure is located on an SiO$_2$ substrate ($\varepsilon_s = 2.25$) and is surrounded by air from the top, as shown in Fig.~\ref{AbsorptionSpectra}a. The lattice has a period of $a = 500$~nm and a layer height of $h = 220$~nm. The blue elements in Fig.~\ref{AbsorptionSpectra}a correspond to an isotropic reciprocal chiral material with the parameters $\varepsilon = 12.25 + 0.01i$, $\mu=1$ and $\edit{\xi = \zeta} = \pm 0.1i$, while the grey elements denote a non-chiral, isotropic reciprocal material characterized by $\varepsilon_m = 2.25 + 0.01i$, $\mu=1$ and $\edit{\xi = \zeta} = 0$.

Using Scheme 2, we calculate the absorption spectra for incident p-polarized light. Figure~\ref{AbsorptionSpectra}b presents the absorption at a fixed in-plane wavevector ($k_1 = 0.1~\mu$m$^{-1}$) for both chiral ($\edit{\xi} = \pm 0.1i$) and non-chiral ($\edit{\xi} = 0$) cases along with their difference. The spectra exhibit pronounced peaks corresponding to \edit{guided mode resonances} (also known as quasiguided modes). One can see that introducing chirality shifts the peak position, resulting in a redshift or blueshift relative to the non-chiral case that depends on the sign of the macroscopic chirality coefficient $\edit{\xi}$.

The dependence of the absorption coefficient on both the photon energy and in-plane wavevector is shown for the non-chiral case ($\edit{\xi = -\zeta} = 0$) in Fig.~\ref{AbsorptionSpectra}c,d. The right panels display the difference in absorption between the two opposite chiralities. This differential map, with red (positive) and blue (negative) colors, clearly reveals the slight energy shift of the resonant modes induced by non-zero chirality.

\section{Conclusion}

In conclusion, we have developed a comprehensive and advanced formulation of the Fourier modal method tailored for the rigorous analysis of two-dimensionally periodic multilayered structures composed of magneto-electric bi-anisotropic materials characterized by macroscopic complex-valued coefficients $\varepsilon$, $\mu$, $\xi$ and $\edit{\zeta}$. Our work generalizes the conventional FMM framework by incorporating arbitrary $3\times 3$ tensors for the macroscopic magneto-electric coefficients, thereby enabling the study of a vast range of periodic structures with chiral and non-reciprocal materials. An important feature of this study is the detailed comparison of two numerical schemes: one implementing generalized Lifeng Li's factorization rules and one without. We have derived explicit expressions for the Fourier tensors in both cases, demonstrating their correct reduction to established forms in the limit of absence of magneto-electric coupling. Crucially, our analysis confirms that the application of factorization rules remains essential, as this scheme delivers superior convergence rates even for structures exhibiting large macroscopic chirality. Therefore, this enhanced formulation establishes itself as a fast, rigorous, and versatile computational technique for the design and theoretical investigation of next-generation photonic devices leveraging the full potential of advanced chiral and bi-anisotropic materials.

\section{Acknowledgement}
This work was supported by the Russian Science Foundation (project 25-12-00454). S.D. acknowledges Maxim Gorlach for a fruitful discussion.

\section*{Appendix A. Analytical solution in a slab with one-dimensional periodicity}
\label{appA}

\begin{figure}[h!]
    \centering
    \includegraphics[width=0.7\linewidth]{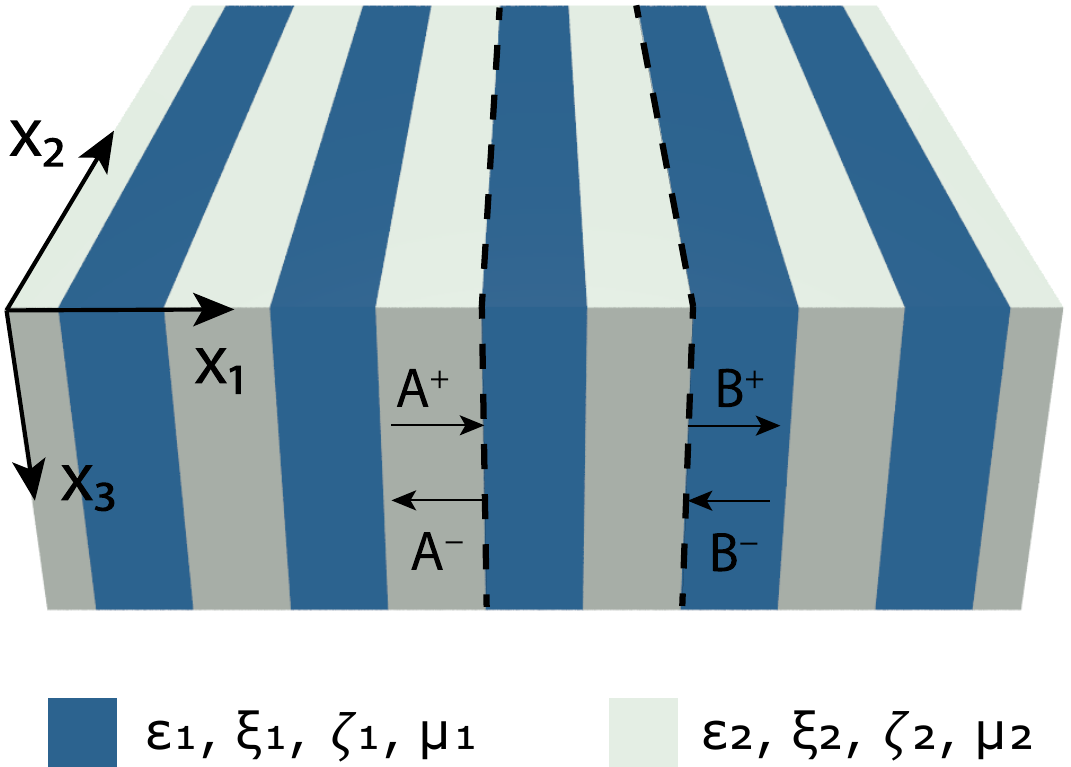}
    \caption{Finding an exact value of $k_3$ in a slab with one-dimensional periodicity. Dashed lines denote boundaries of the considered unit cell.}
    \label{Analytical solution}
\end{figure}

\edit{To date, several methods have been proposed in the literature for solving Maxwell’s equations in bianisotropic media with one-dimensional periodicity~\cite{onishi2011formulation,Lalane}.} To find the exact solution in a periodic slab with one-dimensional periodicity, one can look at it from a different perspective, considering that our slab is a multilayered stratified medium with all layers being homogeneous and infinite in lateral directions. Since we are only interested in finding the exact value $k_3$ in the slab, we do not use its thickness as an input parameter for our problem, nor do we need any information on the \edit{neighboring} layers. Hence, despite the fact that in the FMM all layers (except for the substrate and superstrate) are characterized by a certain thickness, these two representations are fully equivalent for our purpose. Due to homogeneity of the layers in a stratified system, Maxwell's equations can be solved analytically. One of the ways of doing so is using the scattering-matrix formalism, a well established technique for such types of media, since it does not require Fourier transform \cite{lekner1996optical}. 

In order to find the exact value $k_3$ of modes (plane waves) propagating in the stratified medium with an infinite number of periods, we write the definition of the scattering matrix \edit{$\mathcal{S}$} connecting incoming and outgoing amplitudes for one period, as shown in Fig.~\ref{Analytical solution}:
\begin{gather*}
    \edit{\mathcal{S}}(\omega,k_3)
    \begin{pmatrix}
        A^+\\
        B^-
    \end{pmatrix} = 
    \begin{pmatrix}
        B^+\\
        A^-
    \end{pmatrix},
\end{gather*}
where $A^{\pm}$ and $B^{\pm}$ are the amplitudes of plane waves propagating in the positive and negative $x_1$ directions in the stratified periodic medium, and $k_3$ now plays a role of the problem parameter, along with $\omega$. Then, applying Bloch's theorem to the mode's field, 
\begin{gather*}
    B^+ = A^+ e^{ik_1 x_1}, \quad A^- = B^- e^{-ik_1 x_1},
\end{gather*}
we find that plane waves propagating in the stratified periodic medium and characterizing by the photon energy $\omega$ and wavevector $(k_1, 0, k_3)$ must satisfy the following equation:
\begin{gather*}
    \begin{pmatrix}
        e^{-ik_1 x_1} & 0\\
        0 & e^{ik_1 x_1}
    \end{pmatrix}
    \edit{\mathcal{S}}(\omega,k_3)
    \begin{pmatrix}
        A^+\\
        B^-
    \end{pmatrix}
    =
    \begin{pmatrix}
        A^+\\
        B^-
    \end{pmatrix}.
\end{gather*}
Therefore, in order to find the exact value $k_3$ of the modes in the initial periodic slab with one-dimensional periodicity, one should vary $k_3$ as an independent parameter at fixed $\omega$ and $k_1$, to obtain the matrix \edit{$\mathcal{C}$}
\begin{equation}
    \edit{\mathcal{C}}(\omega,k_1,k_3) = \begin{pmatrix}
        e^{-ik_1 x_1} & 0\\
        0 & e^{ik_1 x_1}
    \end{pmatrix}
    \edit{\mathcal{S}}(\omega,k_3)
\end{equation}
such that one of its eigenvalues equals 1. The procedure will provide us with the exact (analytical) value of $k_3$, accurate to machine precision.

% \nocite{*}
\bibliography{refs}

\end{document}